\documentclass[twocolumn,amsmath,amssymb,aps,pra]{revtex4-2}
\usepackage{bm}
\usepackage{amsfonts}
\usepackage{amsmath}
\usepackage{graphicx}
\usepackage{float}
\usepackage{mathtools}
\begin{document} 
\title{A few-parameter non-cylindrical paraxial optical beam described by the modified Bessel function}
\author{Tomasz Rado\.zycki}
\email{t.radozycki@uksw.edu.pl}
\affiliation{Faculty of Mathematics and Natural Sciences, College of Sciences, Institute of Physical Sciences, Cardinal Stefan Wyszy\'nski University, W\'oycickiego 1/3, 01-938 Warsaw, Poland} 

\begin{abstract}
A few-parameter expression for a light beam is found as a solution of the paraxial Helmholtz equation. It is achieved by exploiting appropriately chosen complex variables which entail the separability of the equation. Next, the expression for the beam is obtained independently by superimposing shifted Gaussian beams, whereby the shift can be made either by a real vector (in which case the foci of the Gaussian beams are located on a circle) or by a complex one.  The solutions found depend on several parameters, the specific choice of which allows to obtain beams with quite different properties. For several selected parameter values figures are drawn, demonstrating the spatial distribution of the energy density and phase. In special cases, the effect of a shift of the intensity peak from one branch to another and phase singularities are observed.
\end{abstract}

\maketitle

\section{Introduction}\label{intr}

Within the scalar optics approximation a number of monochromatic paraxial light beams of intriguing properties have been found. From the mathematical perspective, assuming the propagation  along the $z$-axis, they are described  as solutions of the so-called paraxial equation:
\begin{equation}\label{paraxial}
\mathcal{4}_\perp \psi(\bm{r},z)+2ik\partial_z \psi({\bm{r}},z)=0,
\end{equation}
where the stationary function $\psi(\bm{r},z)$, called the {\em envelope},
is related to the electric field via
\begin{equation}
\bm{E}(\bm{r},z,t)=\bm{E}_0 e^{ik(z-ct)}\psi(\bm{r},z),
\label{epi}
\end{equation}
with $\bm{E}_0$ representing a constant vector. 
The Laplace operator $\mathcal{4}_\perp$ in (\ref{paraxial}) is the two-dimensional one acting in the transverse plane only (here $\bm{r}=[x,y]$). The symbol $\partial_z$ stands for the partial derivative $\frac{\partial}{\partial z}$. The approximations leading to the form (\ref{paraxial}) have been discussed in detail elsewhere~\cite{sie,lax}. 

A commonly accepted fundamental solution of the paraxial equation is the Gaussian beam, which in the cylindrical coordinates has the form (apart from the normalization constant):
\begin{eqnarray}
\psi(r,\varphi,z)=&&\;\Big(\frac{w_0}{w(z)}\Big)^{n+1}r^ne^{i n\varphi}\label{gauss}\\
&&\times\exp\Big[-\frac{r^2}{w(z)^2}+i\frac{kr^2}{2 R(z)}
-i(n+1)\psi_G(z)\Big]\nonumber
\end{eqnarray}
endowed ($n\neq 0$) or not ($n=0$) with orbital angular momentum with respect to the propagation axis \cite{kl,davis,sie,nemo,mw,saleh,ibbz,sesh,gustavo,er,selina}. The basic parameters that characterize this beam ($w_0$ -- the waist radius, $w(z)=w_0\sqrt{1+(z/z_R)^2}$ -- radius at a distance $z$,
$R(z)=z(1+(z_R/z)^2)$ -- the wavefront-curvature radius,  $z_R=kw_0^2/2$ -- the Rayleigh length and $\psi_G(z)= \arctan (z/z_R)$ -- the Gouy's phase) are well known and need no further explanation. This beam, written in a slightly modified although equivalent form,  will be of use in the present work (see~(\ref{psig})).

Other beams of similar nature include Bessel-Gaussian (BG) \cite{saleh,gori,april1,mendoza}, modified BG \cite{bagini,li,trsh}, Laguerre-Gaussian (LG)  \cite{sie,saleh,mendoza,lg,lg2,april2,april3,nas}, Kummer-Gaussian (KG) (i.e.,  Hypergeometric-Gaussian)~\cite{kot,karimi}, or $\gamma$ beams \cite{trgam}, the latter containing no Gaussian fall-off factor. All of them are cylindrical in the sense that the wave intensity displays axial symmetry.
Equation (\ref{paraxial}) has also solutions of another kind, which do not manifest cylindrical symmetry. As examples, one can mention Hermite-Gaussian (HG) beams~\cite{sie,praro} of rectangular symmetry, or Airy beams \cite{kalmi,besie,svi}. 

The literature on this subject is extremely vast due to the important and broad applications of light beams, especially those with nontrivial structure, ranging from trapping and guiding of particles, through image processing, optical communication, harmonics generation in nonlinear optics, quantum cryptography, up to biology and medicine.

Some attempts to obtain a more general description and/or derivation of various paraxial beams have been undertaken in the past ~\cite{end,li,vl,fe,qw,trhan}. In this paper, we wish to present, along these lines, somewhat more general solution to the equation~(\ref{paraxial}), which depends on several parameters remaining at our disposal. These few-parameter solution does not exhibit cylindrical symmetry (in the sense spoken of above), except for some special cases. A particular choice of the parameters values enables on one hand to recover some of the above-mentioned modes, and on the other to obtain other modes with equally interesting properties. 

The solution in question will be obtained below in two ways. First, in Section~\ref{der}, some specially defined substitutions are used, which leads in two steps to the complete separation of the paraxial equation. It is known that the Helmholtz equation in $3$ dimensions, owing to the Robertson-Eisenhart condition~\cite{rob,eis}, turns out to be separable in $11$ orthogonal coordinate systems~\cite{sep}. However, in order to obtain new solutions, in this work complex variables will be employed.

Then, in Section~\ref{exp}, it is demonstrated that, at least for integer values of the parameter $l$ (see (\ref{sepn})), these solutions are superpositions of shifted zero-order Gaussian beams~\cite{shift} with some appropriately tailored weight function. It should be stressed that the first approach does not require $l$ to be an integer, thus being more general.  Up to our knowledge such a solution of the paraxial equation has not been published before.

Section~\ref{spca} is devoted to certain special properties of so derived beams. In particular, it is shown how a concrete choice of parameter values ($l,\mu,\chi$) leads to known solutions previously obtained in the literature. Then the properties of the waves of Section~\ref{der} in their general form are analyzed and their spatial distributions are plotted, for several selected parameter values. Necessarily, we had to limit ourselves here to merely few cases, since there are many possible options, especially when considering that the parameters can also assume complex values. In some cases, these spatial distributions exhibit quite special properties, which deserve some attention and are discussed in following sections in detail. Here let us only mention a transfer of the energy-density peak between two beam-forming branches.

\section{Derivation of the few-parameter formula for paraxial beams}\label{der}

In place of Cartesian coordinates $x,y,z$ let us introduce in~(\ref{paraxial}) three complex variables $\xi,\eta,\alpha$ defined in the following way
\begin{subequations}\label{defvar}
\begin{align}
&\xi(x,y,z)=\left(\mu+\frac{2\chi}{\alpha(z)}(x+i y)\right)^{1/2},\label{defvarxi}\\
&\eta(x,y,z)=\left(\mu+\frac{2\chi}{\alpha(z)}(x-i y)\right)^{1/2},\label{defvareta}\\
&\alpha(z)=w_0^2+\frac{2 i z}{k}, \label{defvaralpha}
\end{align}
\end{subequations}
where quantities $\mu$ and $\chi$ are certain parameters. Their values stay at our disposal for the moment and their role in the structure of the beam will be determined later. The choice of the particular branches of the complex roots in~(\ref{defvarxi}) and (\ref{defvareta}) is inessential for the energy density represented, up to a constant, by $|\psi(\bm{r},z)|^2$. As regards the third variable, i.e. $\alpha$, it is in fact the quantity known as the ``complex beam parameter'', apart from some constant coefficient. Consequently $w_0$ stands for the beam's waist radius.

Now let us try to rewrite the paraxial equation in terms of these new variables isolating, however, from the very beginning the standard Gaussian factor. In order to achieve this we set
\begin{equation}\label{isog}
\psi(\bm{r},z)=e^{-\frac{r^2}{\alpha}}\tilde{\psi}(\xi,\eta,\alpha)
\end{equation}
and derive the differential equation satisfied by $\tilde{\psi}$. As can be verified in the straightforward way, this equation takes the form
\begin{equation}\label{eqx}
\frac{1}{\xi\eta}\,\frac{\partial^2 \tilde{\psi}}{\partial\xi\partial\eta}-\frac{\alpha^2}{\chi^2}\,\frac{\partial \tilde{\psi}}{\partial\alpha}-\frac{\alpha}{\chi^2}\,\tilde{\psi}=0.
\end{equation}
The solution of~(\ref{eqx}) can be looked for in the form of the product
\begin{equation}\label{sola}
\tilde{\psi}(\xi,\eta,\alpha)=A(\xi,\eta)B(\alpha),
\end{equation}
which allows for the full separation of the variables $\xi,\eta$ from $\alpha$:
\begin{equation}\label{sepa}
\frac{1}{\xi\eta A(\xi,\eta)}\frac{\partial^2 A}{\partial\xi\partial\eta}=\frac{\alpha}{\chi^2}\left(\frac{\alpha}{B(\alpha)}\,\frac{\partial B}{\partial\alpha}+1\right).
\end{equation}
Both sides of this equation have to be equal to the same constant, which, by virtue of $\chi$ being arbitrary at this point (real or complex), can be set equal to $1$. The solution of the first equation
\begin{equation}\label{eqB}
\frac{\partial B}{\partial\alpha}=\frac{1}{\alpha}\left(\frac{\chi^2}{\alpha}-1\right)B(\alpha),
\end{equation}
can be obtained in the standard way in the form
\begin{equation}\label{solB}
B(\alpha)=B_0\,\frac{1}{\alpha}\,e^{-\frac{\chi^2}{\alpha}},
\end{equation}
with $B_0$ standing for a certain constant. The second equation, i.e.,
\begin{equation}\label{eqA}
\frac{1}{\xi\eta}\frac{\partial^2 A}{\partial\xi\partial\eta}=A(\xi,\eta),
\end{equation}
can be solved as well upon first introducing two new variables
\begin{equation}\label{uv}
u=\frac{\xi}{\eta},\;\;\;\; v=\xi\,\eta,
\end{equation}
and then assuming
\begin{equation}\label{asep}
A(\xi,\eta)=A_u(u)A_v(v).
\end{equation} 
Standard variables-separation procedure leads to two equations
\begin{subequations}\label{sepn}
\begin{align}
&A_u''+\frac{1}{u}\,A_u'-\frac{l^2}{u^2}\,A_u=0,\label{sepnu}\\
&A_v''+\frac{1}{v}\,A_v'-\left(1+\frac{l^2}{v^2}\right)A_v=0,\label{sepnv},
\end{align}
\end{subequations}
with $l^2$ standing for a separation constant. It should be pointed out here that $l$, despite the symbol used, need not be an integer. It can represent a fractional, real or even complex number, and the separation of the variables in the equation~(\ref{eqA}) proceeds in the same manner. This fact implies that the resultant expression~(\ref{famsol}) will describe a whole wide variety of beams of different nature, depending on the choice made for the parameter's value. Several interesting examples will be provided in Sect.~\ref{spca}.

Equation~(\ref{sepnu}) has the two obvious solutions
\begin{equation}\label{ausol}
A_u(u)=u^l,\;\;\;\; A_u(u)=u^{-l},
\end{equation}
but the latter can be omitted as merely leading to the replacement $y\mapsto -y$. In turn Eq.~(\ref{sepnv}) is the modified Bessel equation with a general solution in the form
\begin{equation}\label{ges}
A_v(v)=C_1 I_l(v)+C_2K_l(v),
\end{equation}
where $I_l(v)$ is the modified Bessel function and $K_l(v)$ is the Macdonald (Basset) function. In the present paper we set $C_2=0$ and concentrate on the solution in the form of the modified Bessel function which exhibits certain peculiar properties. Consequently, given the formulas~(\ref{isog}), (\ref{sola}), (\ref{solB}) and (\ref{sepn}), the full family of beams representing the solution of the paraxial equation, labeled with the value of the parameter $l$, becomes
\begin{eqnarray}
\psi_l(\bm{r},z)=&&\frac{N}{\alpha(z)}\left(\frac{\xi(x,y,z)}{\eta(x,y,z)}\right)^lI_l\big(\xi(x,y,z)\eta(x,y,z)\big)\nonumber\\
&& \times e^{\textstyle-\frac{x^2+y^2+\chi^2}{\alpha(z)}}.\label{famsol}
\end{eqnarray}
with $\xi(x,y,z),\eta(x,y,z)$ and $\alpha(z)$ defined in~(\ref{defvar}) and $N$ standing for a certain normalization constant. For the case of $l$ being an integer, in the following section the representation in the form of superimposed shifted Gaussian beams is given.

\section{Expansion onto shifted Gaussian beams}\label{exp}

Consider now the $0$-order Gaussian beam, which constitutes the fundamental solution of the paraxial equation~(\ref{paraxial}). Apart from certain coefficient which is not involved in this work, it has the form
\begin{equation}\label{psig}
\psi_G(\bm{r},z)=\frac{1}{\alpha(z)}\,e^{\textstyle-\frac{x^2+y^2}{\alpha(z)}}.
\end{equation}
Below we will try to construct the beam found in the previous section, for the special case of $l\in \mathbb{Z}$, and expressed through expression~(\ref{famsol}), out of fundamental Gaussian beams whose symmetry axes are no longer the $z$-axis but are shifted from it by some vector $[\chi\cos\beta,\chi\sin\beta]$, i.e., out of
\begin{equation}\label{psigs}
\psi_{G\beta}(\bm{r},z)=\frac{1}{\alpha(z)}\,e^{\textstyle-\frac{(x-\chi\cos\beta)^2+(y-\chi\sin\beta)^2}{\alpha(z)}}.
\end{equation}
Such an expression, however, satisfies the paraxial equation even if $\chi$ is complex, in which case the geometrical interpretation mentioned above is modified, but the expression~(\ref{psigs}) remains effective. Therefore, we do not restrict ourselves to the real case. Various shifted, i.e. off-axis beams, have been dealt with in different context~\cite{shift,kov1,kov2}.

This kind of beams, shifted by real or complex vectors, can be superimposed with some amplitude $f(\beta)$, which can be tailored to one's needs. Below it is assumed to be periodic, with the period of $2\pi$, i.e.,
\begin{equation}\label{fper}
f(\beta+2\pi)=f(\beta).
\end{equation}
Let us then consider the following superposition of $\psi_{G\beta}$'s:
\begin{eqnarray}\label{supg}
\psi(\bm{r},z)&&=\int\limits_0^{2\pi}d\beta f(\beta)\psi_{G\beta}(\bm{r},z)\\
&&= \frac{1}{\alpha(z)}\int\limits_0^{2\pi}d\beta f(\beta)\,e^{\textstyle -\frac{r^2+\chi^2-2\chi r\cos(\beta-\varphi)}{\alpha(z)}}\nonumber\\
&&= \frac{1}{\alpha(z)}\int\limits_0^{2\pi}d\beta f(\beta+\varphi)\,e^{\textstyle -\frac{r^2+\chi^2-2\chi r\cos\beta}{\alpha(z)}},\nonumber
\end{eqnarray}
where polar coordinates ($r$,$\varphi$) are introduced. The last expression is owed to the periodicity of the coefficient function $f(\beta)$.

A variety of periodic functions $f(\beta)$ may be chosen at this point, but for our purposes it is convenient to substitute
\begin{equation}\label{self}
f(\beta)=\frac{1}{2\pi}\, e^{\textstyle i l \beta} e^{\textstyle\mu \cos\beta}.
\end{equation}
The requirement of periodicity entails letting $l$ be an integer, and $\mu$ is an arbitrary (possibly also complex) parameter. If $\mu$ has a non-vanishing real part, then different shifted Gaussian mods~(\ref{psigs}) enter with different weights, whereby the resultant beam no longer exhibits cylindrical symmetry. For purely imaginary $\mu$, there might be no symmetry either due to different phase factors and to the interference. 

Using the weight function in the form
\begin{equation}\label{selfd}
f(\beta)=\frac{1}{2\pi}\, e^{\textstyle i l \beta} e^{\textstyle\mu \cos(\beta-\delta)},
\end{equation}
one can easily rotate the beam by an angle $\delta$.

With this choice of the coefficient function $f(\beta)$, the beam $\psi(\bm{r},z)$ may be given the form
\begin{eqnarray}
\psi_l&&(r,\varphi,z)=\frac{1}{2\pi\alpha(z)}\, e^{\textstyle i l \varphi}e^{\textstyle-\frac{r^2+\chi^2}{\alpha(z)}}\label{dfg}\\
&&\times  \int\limits_0^{2\pi}d\beta\, e^{\textstyle i l\beta}e^{\textstyle(\mu\cos\varphi+\frac{2\chi r}{\alpha(z)})\cos\beta-\mu\sin\varphi\sin\beta}. \nonumber
\end{eqnarray}
The integral with respect to $\beta$ can now be executed according to the formula
\begin{eqnarray}
\int\limits_0^{2\pi}d\beta\, e^{\textstyle i l\beta}&&e^{\textstyle q\cos\beta+p\sin\beta}=2\pi\left(\frac{q+ip}{q-ip}\right)^{l/2}\nonumber\\&&\times  I_l(\sqrt{q^2+p^2}).\label{intbe}
\end{eqnarray}
which holds for arbitrary complex values of $q$ and $p$. Substituting
\begin{subequations}\label{pq}
\begin{align}
&q=\mu\cos\varphi+\frac{2\chi r}{\alpha(z)},\label{pqq}\\
&p=-\mu\sin\varphi,\label{pqp}
\end{align}
\end{subequations}
one finds that
\begin{subequations}\label{qp2}
\begin{align}
&q+ip=e^{\textstyle -i \varphi }\, \xi^2,\label{qp2a}\\
&q-ip=e^{\textstyle i \varphi }\, \eta^2.\label{qp2b}
\end{align}
\end{subequations}
Finally the following formula is obtained
\begin{eqnarray}
\psi_l(\bm{r},z)=\frac{1}{\alpha(z)}&&\,e^{\textstyle-\frac{x^2+y^2+\chi^2}{\alpha(z)}}\left(\frac{\xi(x,y,z)}{\eta(x,y,z)}\right)^l\nonumber\\
&& \times I_l\big(\xi(x,y,z)\eta(x,y,z)\big),\label{famsoli}
\end{eqnarray}
which is identical to~(\ref{famsol}) apart from the normalization constant. It is obvious that this expression does satisfy the paraxial equation as each of the terms in the linear superposition does.

\section{Beams' properties}\label{spca}

\subsection{General remarks} \label{gere}

It is clear that the energy density of the wave integrated in any perpendicular plane $z=\mathrm{const}$, and  proportional to
\begin{equation}
\int \mathrm{d}^2r |\psi(\bm{r},z)|^2,
\label{norm}
\end{equation}
is finite owed to the presence of the Gaussian factor. It ensures the convergence of~(\ref{norm}) and dictates the asymptotic behavior of the beam away from the optical axis despite the eventual divergence of the Bessel function (depending on the value of $\chi$) for large $r$:
\begin{equation}\label{abe}
I_l\big(\xi\eta\big)\sim e^{\textstyle\frac{2\chi}{\alpha}\,r}.
\end{equation}

It is also obvious that the value of the integral~(\ref{norm}) is constant (i.e., it does not depend on $z$). If $z$ were treated as time variable this would correspond to the conservation of the quantum-mechanical probability in the evolution governed by the two-dimensional Schr\"odinger equation. 

The  family of beams introduced in this paper, apart from the beam waist $w_0$, includes a couple of parameters,  the values of which stay at our disposal, as $l$, $\mu$ and $\chi$. According to the results of Section~\ref{der}, there are no limitations for their values. $l$ can be either integer or fractional as well as real or complex. The same refers to $\mu$ and $\chi$. Consequently the general structure of~(\ref{famsol}) is potentially very rich. For a few special choices, the general expression~(\ref{famsol}) gets reduced to well-known types of beams of different properties. For instance setting $\chi=0$ the fundamental Gaussian beam~(\ref{psig}) is obtained. One then has
\begin{equation}
\frac{\xi}{\eta}=1,\;\;\;\; \mathrm{and}\;\;\;\; \xi\eta=\mu.
\label{cpxe}
\end{equation}
Consequently the Bessel function in~(\ref{famsol}) reduces to an overall multiplicative constant similarly as the integration with respect to $\beta$. Only Gaussian factor (together with $1/\alpha$) survives. This is an obvious result since the shifting vector in (\ref{psigs}) gets null in this case.

If the value of the second parameter (i.e., $\mu$) equals zero (surely, now with $\chi\neq 0$), the beam recovers its cylindrical character. One then gets
\begin{equation}
\frac{\xi}{\eta}=e^{i\varphi},\;\;\;\; \mathrm{and}\;\;\;\; \xi\eta=\frac{2\chi}{\alpha}\,r.
\label{upxe}
\end{equation}
In this case for the real value of the parameter $\chi$, the so-called modified Bessel-Gaussian beam is obtained
\begin{equation}
\psi_{mBG}(r,\varphi,z)=\frac{1}{\alpha(z)}\,e^{\textstyle-\frac{r^2+\chi^2}{\alpha(z)}}e^{\textstyle il\varphi}I_l\Big(\frac{2\chi r}{\alpha(z)}\Big).
\label{mgaussbebeam}
\end{equation}
and the imaginary value of $\chi$ leads to the ordinary Bessel-Gaussian beam
\begin{equation}
\psi_{BG}(r,\varphi,z)=\frac{1}{\alpha(z)}\,e^{\textstyle-\frac{r^2-|\chi|^2}{\alpha(z)}}e^{\textstyle il\varphi}J_l\Big(\frac{2|\chi| r}{\alpha(z)}\Big).
\label{gaussbebeam}
\end{equation}
In this case parameter $\chi$ corresponds to $z_R\sin\theta$, where $\theta$ is the half-aperture angle~\cite{bor,mad}.

In general, the parameter $\mu$ to some extent accounts for the asymmetrical character of the beam's intensity. As can be seen from the expression~(\ref{famsol}) and will be shown in the figures below, beams of the class dealt with here in general do not exhibit the cylindrical symmetry. For $\mathrm{Re}\,\mu\neq 0$ the weight function $f(\beta)$ is non-uniformly distributed over the circle, so different shifted Gaussian beams enter with various weights. But also for purely imaginary values of $\mu$ the asymmetry associated with different phase factors and with interference of waves does emerge. As $\mu$ approaches $0$ the beam becomes more and more axially symmetric. The same is observed for the limit of large $\mu$ (more precisely for $|\mu|\gg|\chi|/w_0$) since in this case the deviations occur only far from the $z$ axis, i.e in the region which is irrelevant, as it remains virtually off-beam due to Gaussian damping.

The properties discussed above show that the conditions of the paraxial approximation for the derived beam should be fulfilled no worse as for standard beams: Gaussian and Bessel-Gaussian (regular and modified) for the same values of parameters $\chi$ and $w_0$. A significant role is played here by the value of the parameter $\gamma:=\chi/\mu$, which in some sense `interpolates' between the former and the latter beams. In particular, as the above analysis demonstrates, a small value of $\gamma$ yields a Gaussian beam and its large value a BG or a mBG beam depending on whether $\chi$ is real or imaginary, so the conditions of the paraxial approximation should be met equally well in the present case.  These conclusions are also confirmed by the construction presented in Sec.~\ref{exp}. It is apparent that, for $\chi\in \mathbb{R}$, the solution (\ref{famsol}) is a superposition of shifted Gaussian beams with foci distributed over a circle just as happens for mBG beams~\cite{bagini}. The modification is merely the choice of a non-trivial weighting function (\ref{self}). The same thing is true for imaginary $\chi$: in this case one has to do with a superposition of inclined Gaussian beams with foci at the origin of the coordinate frame, which for a constant weighting function would lead to the formation of a regular BG beam \cite{bagini}.

In the following, the intensity and phase distributions of the beams~(\ref{famsol}) in space will be graphically represented for chosen values of the parameter $l$. It should be stressed that the illustrations presented below are merely examples, since, due to the presence of several parameters there is a number of possible choices of their values leading to different beams' shapes. The specific data are picked so that certain characteristic features be clearly visible in the figures. Additionally, we restrict ourselves in this paper to real values of $\chi$.

\begin{figure*}[!]
\begin{center}
\includegraphics[width=1\textwidth,angle=0]{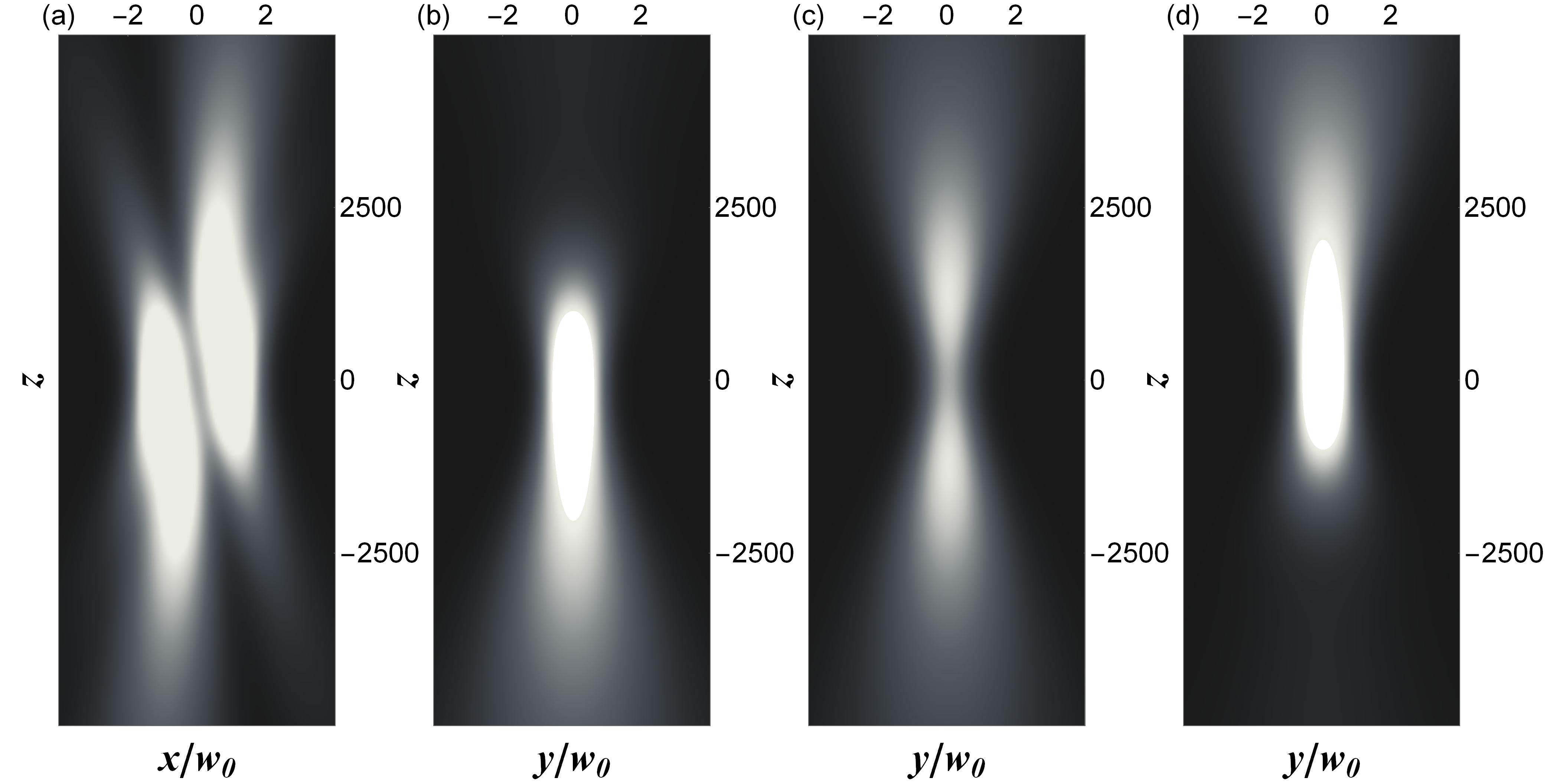}
\end{center}
\vspace{-4.5ex}
\caption{The energy density of the beam~(\ref{famsol}) for $l=1$, $w_0=50$, $\chi=50$ and $\mu=25i$. Plot (a) is performed in the plane $y=0$, and three subsequent plots, (b), (c) and (d) show the beam in the planes $x=-w_0,0,w_0$ respectively. The units on the axes are dimensionless, as they refer to the system in which $k=1$. Bright regions represent high wave intensity and dark region low one.}
\label{intlint}
\end{figure*}

\begin{figure*}[!]
\begin{center}
\includegraphics[width=1\textwidth,angle=0]{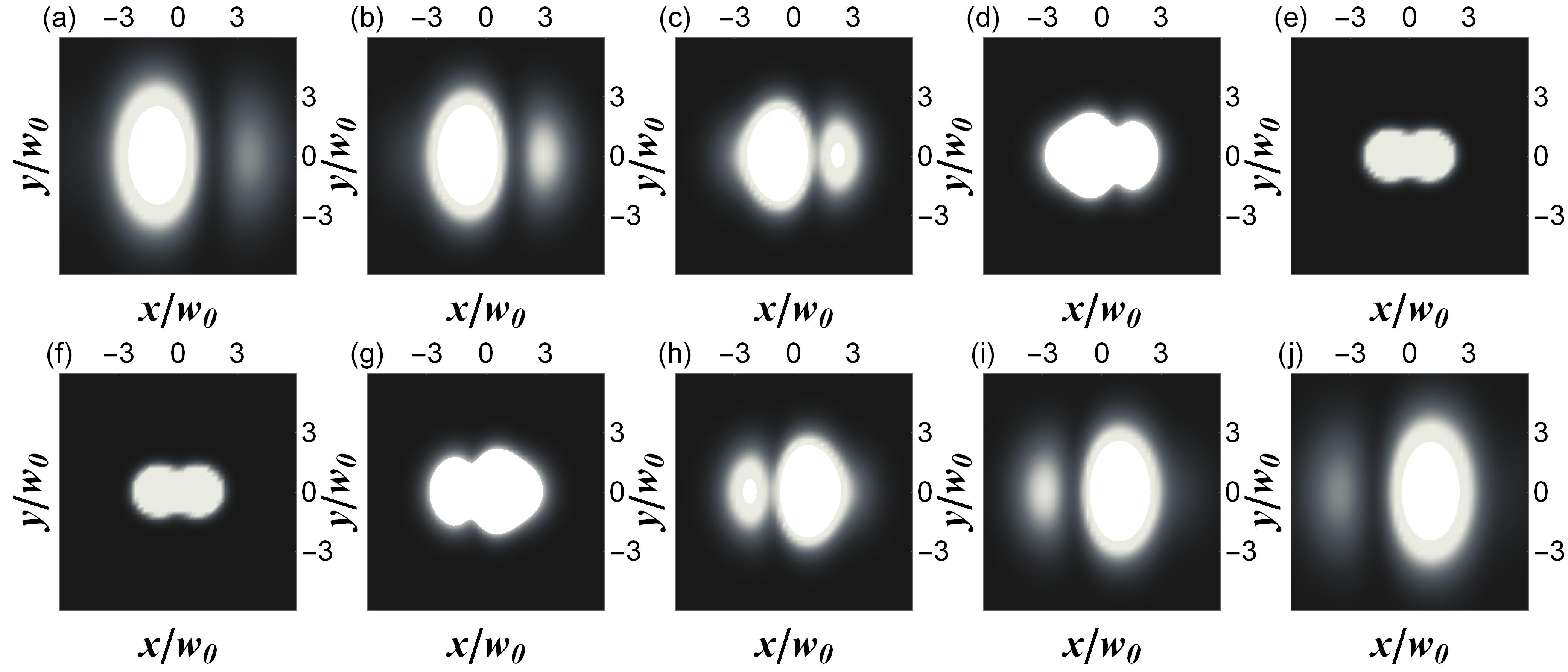}
\end{center}
\vspace{-4.5ex}
\caption{Same as Fig.~\ref{intlint}, but in the perpendicular planes: (a) $z=-5000$, (b) $z=-4000$, (c) $z=-3000$, (d) $z=-2000$, (e) $z=0$ (i.e., upper row) and   (f) $z=0$, (g) $z=2000$, (h) $z=3000$, (i) $z=4000$, (j) $z=5000$ (i.e., lower row).}
\label{intlcut}
\end{figure*}

\subsection{Integer value of $l$}\label{intv}

In Fig.~\ref{intlint} the beam's energy density for the integer value of $l$ (it has been chosen $l=1$) and imaginary value of $\mu$ is plotted. The first diagram shows the distribution of energy density in the plane $y=0$. The other three present the same quantity in the planes: $x=-w_0$, $x=0$ and $x=w_0$.  

For the data specified in the figure the beam is obviously not axially symmetric. Since all the constituent Gaussian beams located on the circle enter with identical weights, so this effect stems from the different phase factors and from the interference between superimposed waves.

A characteristic feature visible in the figure at first glance is the transfer of the wave-intensity peak from the left to the right branch of the beam after crossing the $z=0$ plane. The same effect is also shown in Figure~\ref{intlcut} in the form of the sequence of intersections with planes perpendicular to the propagation axis $z$. This will be apparent also in later figures performed for a fractional value of $l$. Mathematically, this effect is attributable to the change in the sign of the imaginary part of $\alpha(z)$ and to the property of Bessel functions of complex argument, which, so to speak, ``interpolates'' between $I_l$ and $J_l$.

\begin{figure}[!]
\begin{center}
\includegraphics[width=0.5\textwidth,angle=0]{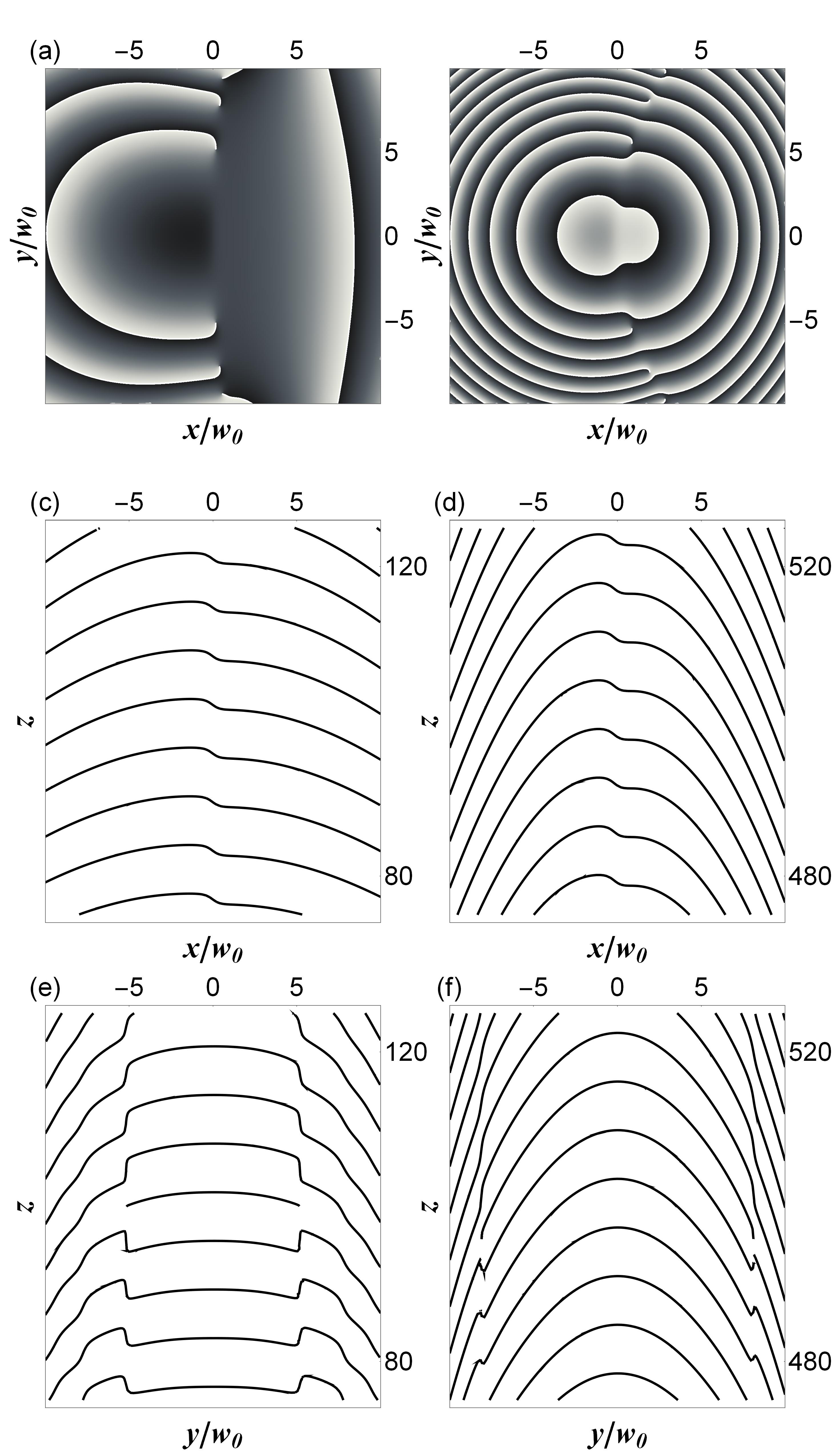}
\end{center}
\vspace{-4.5ex}
\caption{The phases of the beam~(\ref{famsol}), depicted in cut planes: (a) $z=100$, (b) $z=500$. Other parameters are identical as in Fig.~\ref{intlint}. The value of the phase, modulo $2\pi$, is represented continuously by means of the grayscale from $-\pi$ (black color) to $\pi$ (white color). The plots (c) and (d) show the intersection of wavefronts with the plane $y=0$, and (e) and (f) with the planes $x=0.17w_0$, and $x=1.9w_0$ respectively. The latter values are chosen to clearly visualize phase singularities seen in plots (a) and (b). The wavefront surfaces are drawn for integer values of $\pi$.}
\label{intlfullph}
\end{figure}

Physically, the choice of relative phases in the form of the weight function $f(\beta)$ and the values of the parameters (particularly $\mu$) yields a destructive interference on one side and a constructive one on the other.

The spatial energy distribution does not exhibit axial symmetry, as said before, but nonetheless two symmetries can be observed, when the following reflections are carried out:
\begin{enumerate}
\item $y\;\longmapsto \; -y$,
\item $( x,z)\;\longmapsto \; (-x,-z)$.
\end{enumerate}

The former is relatively easy to deduce from~(\ref{famsol}) and~(\ref{defvar}). The product $\xi\eta$ is invariant under this reflection, and the quotient $\xi/\eta$ is insensitive to it provided $y\ll w_0^2 |\mu|/\chi=25 w_0$ (for the data of Fig.~\ref{intlint}), i.e., practically within the entire beam whose transverse size is determined by a Gaussian damping factor. 

The latter symmetry holds in the same sense: both $|\xi/\eta|$ and $|\xi\eta|$ become invariant under simultaneous reflection of $x$ and $z$ if the above condition for $y$ is met. This symmetry is, however, broken for real values of $\mu$ (see Sec.~\ref{fracv}). In this case the shifted Gaussian beams occur in the superposition with significantly different weights (and not just different phase factors), resulting in the loss of some symmetries.

Naturally these conclusions can be deduced from the form of $f(\beta)$ and integral~(\ref{dfg}) as well. For example, a simple substitution $\beta\;\longmapsto\; 2\pi-\beta$ under the integral reveals the aforementioned symmetry with respect to reflection $y\;\longmapsto \; -y$. 

It is clear from these formulas that switching the sign of the parameter $\chi$ merely induces the inversion of the intensity pattern in the plane perpendicular to the propagation axis.
There is not space here to provide further diagrams for larger values of integer $l$, so just note that the beam then looks more and more like a slanted one from paper~\cite{qw}.

In Fig.~\ref{intlfullph}, the phases of the wave-function $e^{ikz}\psi(\bm{r},z)$ are represented in various cut planes. The first two drawings depict in grayscale the distribution of the phase in perpendicular planes (i.e., for $z=\mathrm{const}$). The subsequent four illustrate the profiles of the constant-phase surfaces in different planes parallel to the $z$-axis and at different distances from the focal plane.

In the first two and the last two diagrams, phase singularities resulting from the interference of shifted Gaussian beams are visible. Note the truncated lines of constant phase in drawings (e) and (f) and the concomitant lines' jump, which correspond to the singularities seen in (a) and (c). The values $x=0.17w_0$ and $x=1.9w_0$ were tailored in such a way that the plane of intersection passes through them.

\begin{figure*}[!]
\begin{center}
\includegraphics[width=1\textwidth,angle=0]{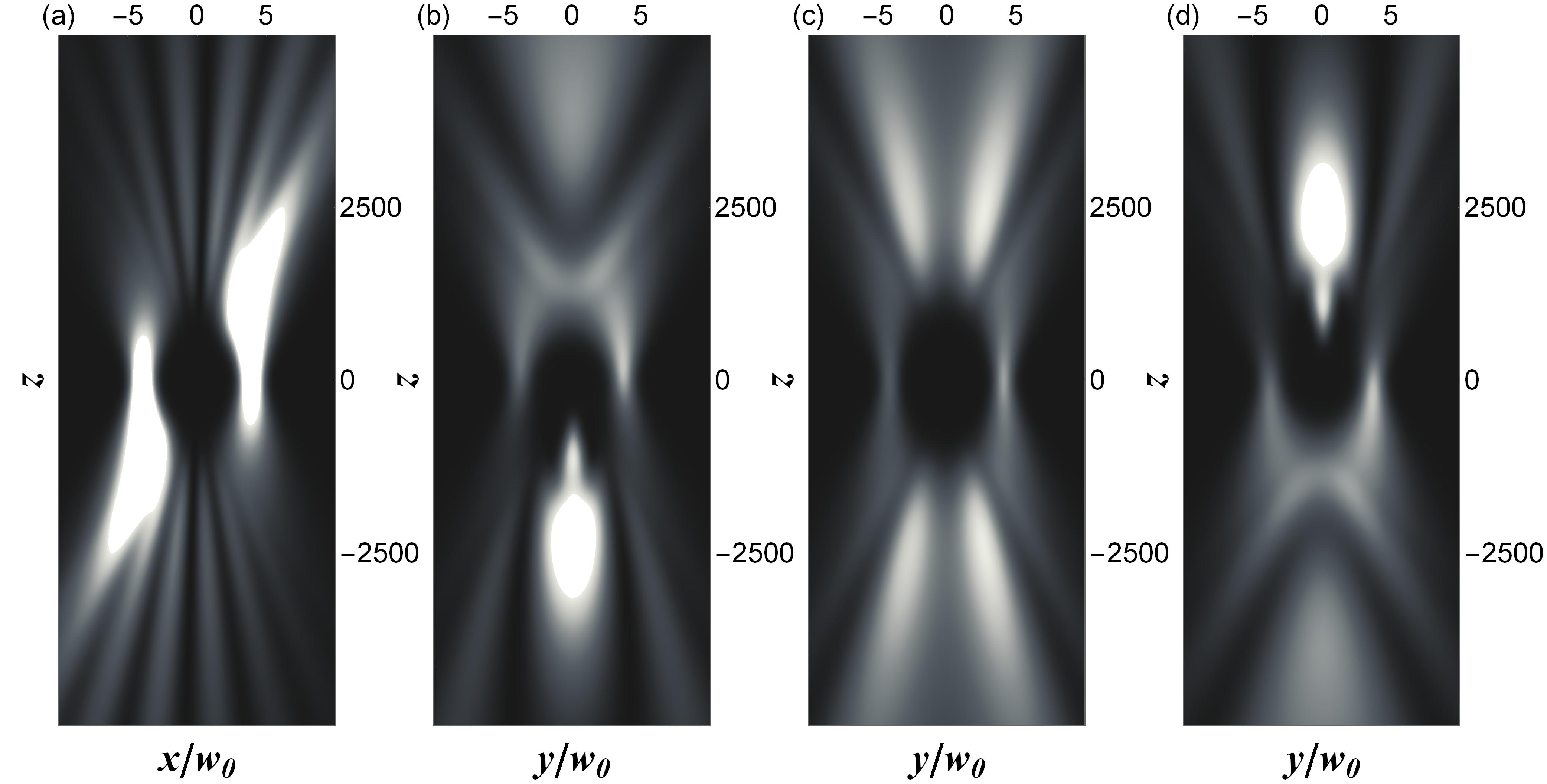}
\end{center}
\vspace{-4.5ex}
\caption{Same as Fig.~\ref{intlint}, but for $l=1/2, w_0=30,\chi=120, \mu=10 i$. The cut planes are: (a) $y=0$, (b) $x=-1.5w_0$, (c) $x=0$ and (d) $y=1.5w_0$.}
\label{fralint}
\end{figure*}

\begin{figure*}[!]
\begin{center}
\includegraphics[width=1\textwidth,angle=0]{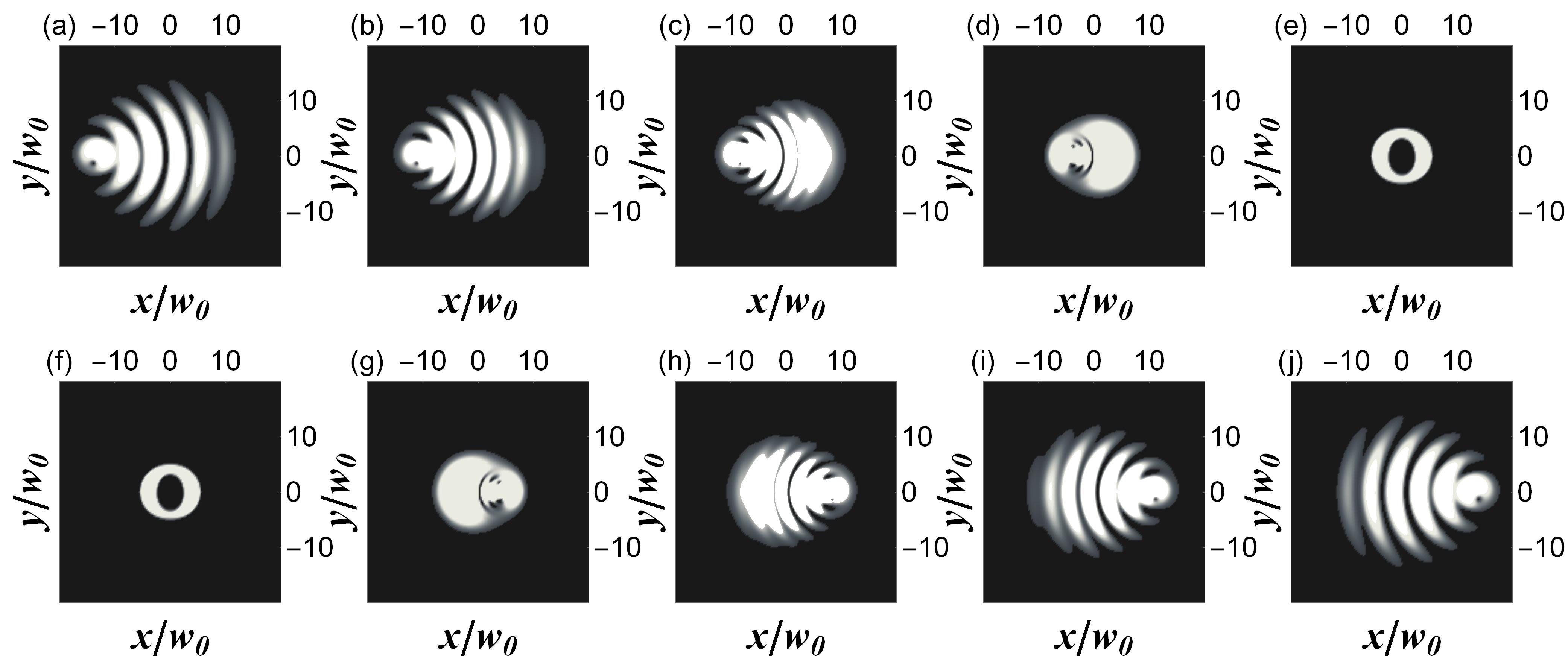}
\end{center}
\vspace{-4.5ex}
\caption{Same as Fig.~\ref{intlcut}, but for the values of parameters of Fig.~\ref{fralint} and for the cut planes: (a) $z=-5000$, (b) $z=-4000$, (c) $z=-3000$, (d) $z=-1500$, (e) $z=0$ (i.e., upper row) and   (f) $z=0$, (g) $z=1500$, (h) $z=3000$, (i) $z=4000$, (j) $z=5000$ (i.e., lower row).}
\label{fralcut}
\end{figure*}

The well-known formula for the generating function of the modified Bessel functions~\cite{span}
\begin{equation}\label{gener}
e^{\textstyle\frac{z}{2}(t+\frac{1}{t})}=\sum_{l=-\infty}^\infty t^lI_l(z),
\end{equation}
allows to construct arbitrary Gaussian beams out of~(\ref{famsol}). Let us consider the sum
\begin{equation}\label{suco}
\sum_{l=-\infty}^\infty a^l \psi_l(\bm{r},z)=\frac{1}{\alpha(z)}\,e^{\textstyle-\frac{x^2+y^2+\chi^2}{\alpha(z)}}\sum_{l=-\infty}^\infty a^l\left(\frac{\xi}{\eta}\right)^lI_l(\xi\eta)
\end{equation}
with certain constant $a$. Applying~(\ref{gener}) one easily gets
\begin{equation}\label{sucol}
\sum_{l=-\infty}^\infty a^l \psi_l(\bm{r},z)=\frac{1}{\alpha(z)}\,e^{\textstyle-\frac{x^2+y^2+\chi^2}{\alpha(z)}}\,e^{\textstyle\frac{1}{2}(a\xi^2+\frac{1}{a}\eta^2)}.
\end{equation}
A simple substitution $a=e^{-i\beta}$ reconstructs now the shifted Gaussian beam~(\ref{psigs}), apart from an overall constant.

\subsection{Fractional value of $l$}\label{fracv}

As an example of a solution with a fractional value of $l$, the value $l=1/2$ was chosen, still with a real value of the parameter $\chi$ and an imaginary value of  $\mu$. The precise values of these and other parameters are again picked out for visibility of plots and are given in the caption of Figure~\ref{fralint}.

For fractional values of $l$, the beam displays a somewhat more complex structure. In particular, it splits into a system of asymmetric rings (in a perpendicular plane), or rather ring segments. This is clearly visible in the Figures~\ref{fralint} and~\ref{fralcut}. Mathematically, for example for $l=1/2$, this follows simply from the fact that the Bessel function $I_{1/2}(z)$ is proportional to $\sinh(z)$ (for other half-values $\cos z$ occurs as well), and therefore for a complex $z=x+iy$
\begin{equation}\label{symi}
I_{1/2}(x+iy)\sim \sinh x\cos y+i\cosh x \sin y.
\end{equation}
The appearance of trigonometric components introduces oscillatory character to the function $\psi(\bm{r}, z)$ resulting in the formation of bright rings.  The centers of these rings follow the two branches of the beam, similar to those of Figure~\ref{intlint}. Identically to the case of $l=1$, the intensity peak is also shifted from one branch to the other.
The observed pattern of the energy density distribution, which is in some aspects similar to the ``asymmetric Bessel modes'' of~\cite{kot1,kot2}, is also repeated for other fractional values of $l$, for which graphic illustration is not provided in this work due to limited space. However, it should be noted that, contrary to Bessel or Bessel-Gaussian beams, there are {\em two} families of ring segments here: one for the left and the other for the right branch. However, the symmetries of the energy density distribution found for integer $l$ still hold.

\begin{figure}[!]
\begin{center}
\includegraphics[width=0.5\textwidth,angle=0]{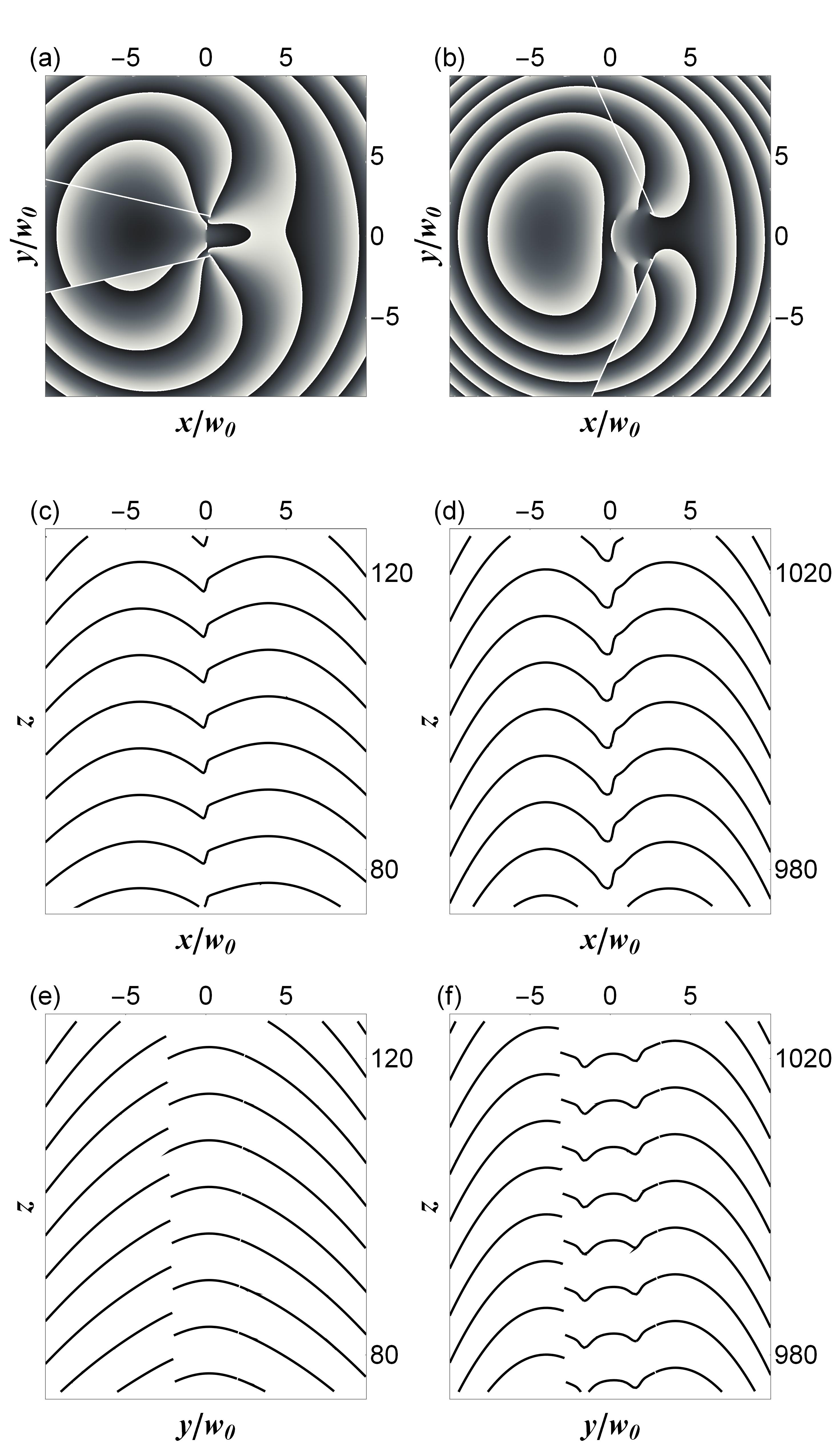}
\end{center}
\vspace{-4.5ex}
\caption{Same as Fig.~\ref{intlfullph}, but for the values of parameters of Fig.~\ref{fralint}, and the for the cut planes: (a) $z=100$, (b) $z=1000$, (c) $y=0$, (d) $y=0$, (e) $x=-4w_0$, (f) $x=2w_0$.}
\label{fralfullph}
\end{figure}

The phase pattern depicted in Figure~\ref{fralfullph} in various cut planes again shows the singularities, in plots (a) and (b) marked with white lines and evidenced by the discontinuities in plots (e) and (f) (the right discontinuity is hardly visible, reducing simply to a white point on the curve due to the jump by $2\pi$). 
 
\begin{figure*}[!]
\begin{center}
\includegraphics[width=1\textwidth,angle=0]{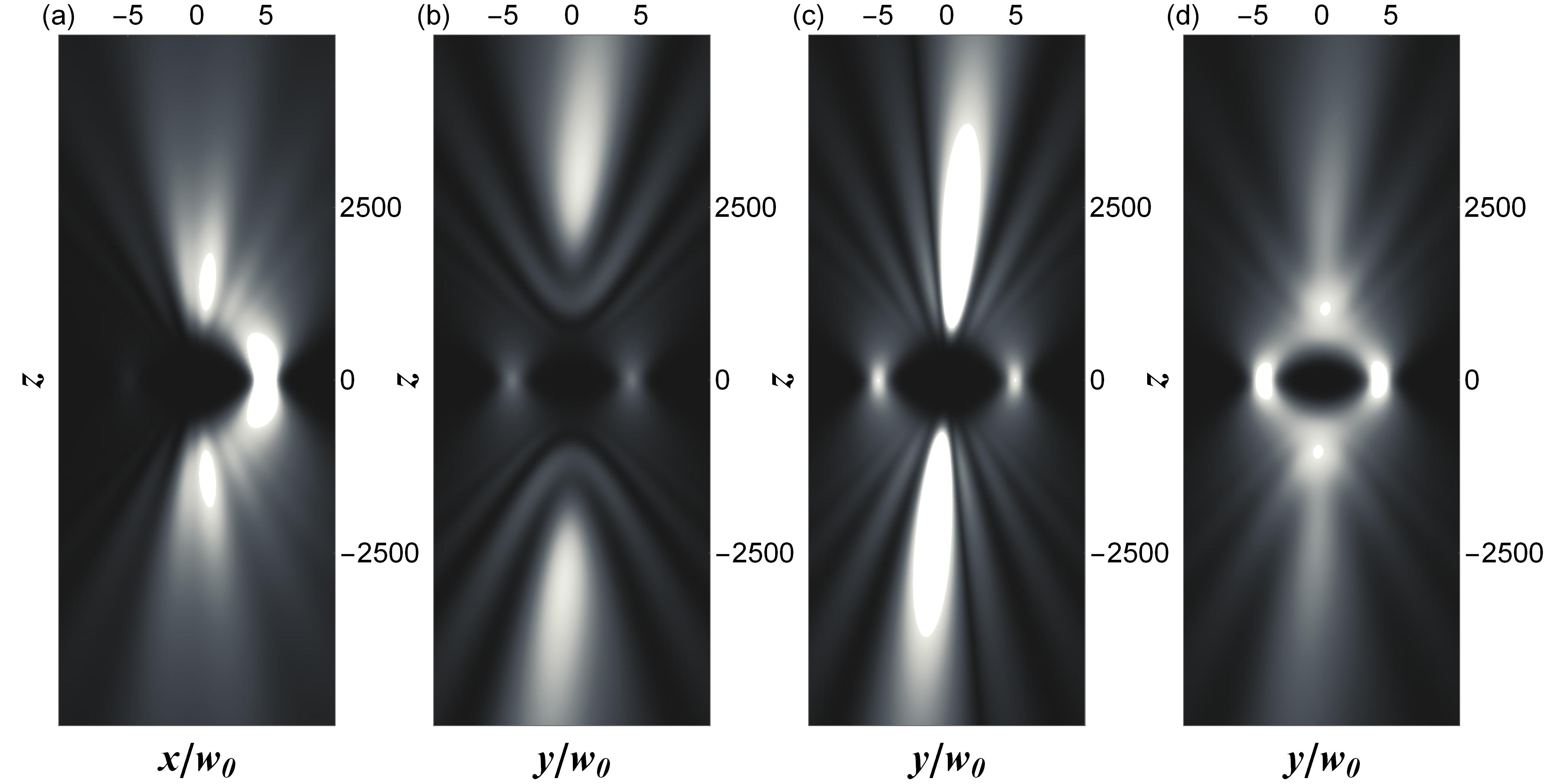}
\end{center}
\vspace{-4.5ex}
\caption{Same as Fig.~\ref{intlint}, but for $l=1/2, w_0=20,\chi=100, \mu=1$. The cut planes are: (a) $y=0$, (b) $x=-2.5w_0$, (c) $x=0$ and (d) $y=2.5w_0$.}
\label{fralint2}
\end{figure*}

\begin{figure*}[!]
\begin{center}
\includegraphics[width=1\textwidth,angle=0]{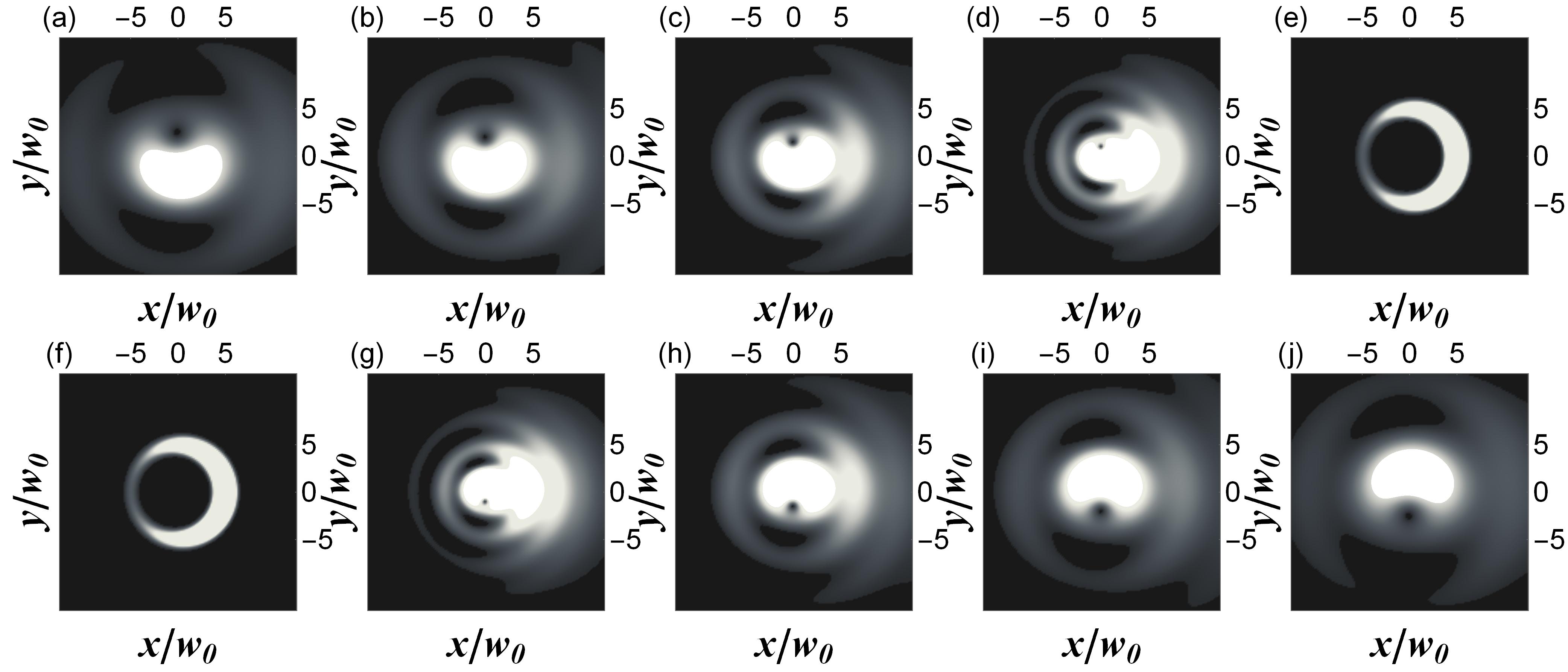}
\end{center}
\vspace{-4.5ex}
\caption{Same as Fig.~\ref{intlcut}, but for the values of parameters of Fig.~\ref{fralint2} and for the cut planes: (a) $z=-5000$, (b) $z=-4000$, (c) $z=-3000$, (d) $z=-2000$, (e) $z=0$ (i.e., upper row) and   (f) $z=0$, (g) $z=2000$, (h) $z=3000$, (i) $z=4000$, (j) $z=5000$ (i.e., lower row).}
\label{fralcut2}
\end{figure*}

As already indicated, in the case of a real value of $\mu$ (assume that $\mu>0$) the weight function $f(\beta)$ does not reduce to a phase factor. Gaussian beams $\psi_{G\beta}(\bm{r},z)$ located in a symmetric way with respect to the $yz$-plane enter with different weights (e.g. $w$ and $1/w$). The beams contributing with larger values of the weight function are those with $-\pi/2<\beta<\pi/2$ for which $\cos\beta>0$. This entails the significantly higher intensity on the right side and the observed energy-density asymmetry. In contrast, the symmetry with respect to the $xz$-plane is maintained due to the invariance of the cosine function when replacing the argument $\beta$ with $2\pi-\beta$. As a result, the beam is still symmetric when transforming $z\;\longmapsto\; -z$ or $y\;\longmapsto\; -y$. In turn, inverting the sign of the parameter $\mu$ (or $\chi$) implies {\em de facto} the reflection of the intensity distribution with respect to the $yz$-plane. This is illustrated in Figures~\ref{fralint2} and~\ref{fralcut2}.

One should still note the dark hole in the middle. It appears for relatively large values of the parameter $\chi$ (such as that chosen in the figures). This effect is quite easy to understand if one remembers that superimposed shifted Gaussian beams are localized on the circle of radius $\chi$. Since its circumference is proportional to $\chi$ and the amplitude of each beam decreases with $\chi$ as $\exp (-\chi^2/w_0^2)$, so roughly speaking and disregarding interference effects, the intensity at the center is proportional to $\chi^2\exp (-2\chi^2/w_0^2)$.
For large values of $\chi$ it becomes extremely low and the resulting beam has a ``holey'' appearance.

\begin{figure}[!]
\begin{center}
\includegraphics[width=0.5\textwidth,angle=0]{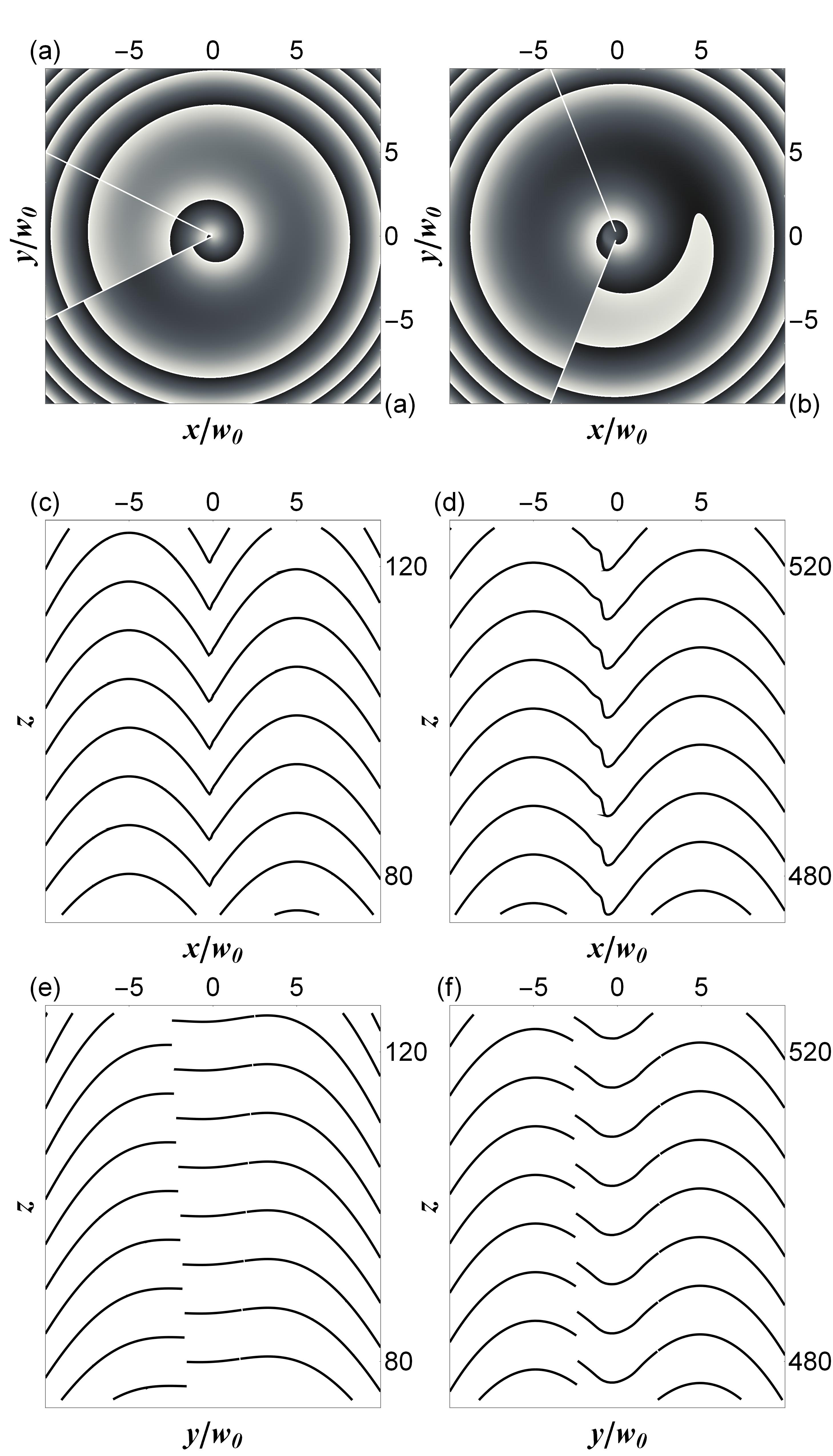}
\end{center}
\vspace{-4.5ex}
\caption{Same as Fig.~\ref{intlfullph}, but for the values of parameters of Fig.~\ref{fralint2}, and the for the cut planes: (a) $z=100$, (b) $z=500$, (c) $y=0$, (d) $y=0$, (e) $x=-4w_0$, (f) $x=-w_0$.}
\label{fralfullphb}
\end{figure}

\section{Summary}\label{sum}

A new solution of the paraxial Helmholtz equation obtained by introducing three complex variables (\ref{defvar}) is provided in this paper. In these variables, the equation was successfully separated.  The obtained solution describes a whole spectrum of light beams with different properties, depending on the choice of values of several parameters.

For integer values of the parameter $l$, the resulting beam was independently constructed as a superposition of shifted Gaussian beams (with foci distributed on a circle) by appropriate selection of the weight function.

Apart from some specific parameter values for which previously known beams (Gaussian, Bessel-Gaussian, modified Bessel-Gaussian) are obtained from the general formula (\ref{famsol}), the spatial distribution of the energy density exhibits certain new properties. In particular, a characteristic feature is the existence of two ``branches'' between which the irradiance transfers in the course of wave propagation. Other peculiarities (such as the appearance of two families of rings, etc.) have been shown in the figures. The phase distribution of the derived waves generally presents a rather intricate pattern that also includes singularities at certain specific locations.

Due to the numerous applications of structured light, it seems that the theoretical derivation of its interesting new forms, which are supposed to be practically realizable (e.g. by appropriately illuminating a computer-generated hologram or computer-controlled spatial light modulator), seems to be of a certain value.  At the same time, it should be emphasized that (\ref{famsol}) offers considerable scope for modification due to the presence of several parameters at the experimentalist's disposal. This property might be for instance used to model the assymmetries of the real aberrated beams with defects caused by experimental setup. An attempt to account for asymmetries in the perturbative way and to decompose the distortion in the base of cylindrical LG modes was proposed in~\cite{lf}. The beam found in this paper has an already built-in asymmetry that can be controlled through accessible parameters, so it seems that it might be more feasible than the use of strictly cylindrical beams.

In conclusion, it can be said that by properly choosing the values of the parameters, this beam can be made closer to real experimental situations than idealized beams showing beautiful mathematical symmetries.

\end{document}